# Smart Massive MIMO: An Infrastructure toward 5th Generation Smart Cities Network

Ahmad Abboud[1], Jean-Pierre Cances[2], Vahid Meghdadi[3], Ali Jaber[4]

[1,2,3] Department Of Components Circuits Signals And High Frequency Systems
Xlim labs, University Of Limoges Limoges, France
[4] Department of Statistic, Lebanese University, Lebanon

*Abstract :* On the Optimizing of Wireless Networks and toward improving the future 5th Generation mobile Network Infrastructure, we propose a novel infrastructure that can be the next Smart City Network.
Our proposed Infrastructure takes into consideration most future demands and challenges, includes Capacity, Reliability, Scalability, and Flexibility. To deal with this issues we propose a wireless network infrastructure that is based on latest technologies of Massive MIMO systems. We further extend our infrastructure with many smart features, to be capable of coping with Cloud Computing, Smartphones, IoT and other intelligence-based services.
The proposed infrastructure uses Network Functions Virtualization (NFV), Software-Defined Networking (SDN), Virtual Antenna Arrays (VAA) and Joint Beamforming to afford flexibility. We further propose a Terminal-centric rather than a Cell-centric based Infrastructure, which optimize interference aware environment and lead to higher capacity and reliability. The new infrastructure includes multi-purpose nodes that run a Network Operating System (NOS). This node will afford a scalable and flexible cost effective and semi-distributed network resources. Other propositions that meet Power-Effective, Cost-Effective, and Scenery aware design are discussed.

## 1 INTRODUCTION

### 1.1 5th Generation Demanding Features

By the revolution of Cloud Computing and Smartphones, a huge amount of data has to traverse the network every second and most of this data are delivered by internet services. The fast growth in bandwidth and QoS demands makes 4th G mobile networks insufficient.

The next generation network must be capable of affording a bandwidth from 100Mbps up to 1Gbps per User Terminal (UT), with a connection density that exceeds 1M connection/Km2 , the mobility of high-speed vehicles up to 500 km/hr and an End to End (E2E) delay less than 10ms.

The 5G network needs to be flexible to be able to afford Network-as-a-Service (NaaS) and easy software defined networking control. Security, Power effective and Cost-effective Services must be afforded. Network Intelligence (NI), Quality of Experience (QoE) are to be included.

### 1.2 SMMIMO and Ongoing Global Research on 5th G infrastructure

Many projects on 5G are ongoing, but we will introduce a small collection that we (from our point of view) consider them the most important. In Europe, the main leading projects are 5G Private Public Partnership (5G PPP) and the 5G Innovation Centre (5G IC). China introduces a primary leading project by IMT-2020 (5G) Promotion Group. Intel Strategic Research Alliance (ISRA) in the USA leads new research on 5G networks.

SoftNet, a software defined decentralized mobile network architecture [1] proposes a dynamically defined architecture, decentralized mobility management, distributed data forwarding, and multi-RATs coordination. The features afforded by SoftNet provides a flexible and high capacity network compared to 4G networks.

Cloud Radio Access Networks (C-RAN), provides low capital and operating expenditures, as well as high spectral efficiency (SE) and energy efficiency (EE).C-RAN decouples the baseband processing from the radio units, allowing the processing power to be pooled at a central location thus reducing the required redundancy. This done by splitting BBU and RRH, where BBU pool offered as a centralized cloud service and RRH's are distributed among the network ends. This architecture solves many challenges face network flexibility but bring a new challenge by flooding front haul links with signaling data to reach BBU resources [2]. Recent research on C-RAN virtualization reduces signaling overhead by logically grouping macro cells with collocated small cells that can provide the core network with a simplified overview [3]. C-RAN projects have been initiated in many organizations such as the European Commission's Seventh Framework Programme and the Next Generation Mobile Networks (NGMN) project.

SoftRAN [4] is a software defined centralized control plane for radio access networks that abstract all base stations in a local geographical area as a virtual big base station comprised of a central controller and radio elements. SoftRAN proposes a logically centralized entity which makes control plane decisions for all the radio elements in the geographical area. However, this architecture still lake some flexibility, where UT's deals with the virtual big base station as a separate BS's. SoftRAN deals with this issue by applying multiple handovers between the same pair of base stations.

### 1.3 SMMIMO Infrastructure Features

This paper proposes a Smart Massive MIMO (SMMIMO) infrastructure as a solution to deal with 5G demands. SMMIMO infrastructure is a hieratical model that connects virtual nodes with each other. Virtual Nodes (VN's) are formed by the cooperation of several resources include (antennas, processing units, memories input output interfaces) distributed on different physical nodes. The Physical Node (PN) can be seen as Base Stations (BS) in classical cellular systems, wherein our model Physical Nodes are base stations that hold a massive number of antennas with high flexible computational and communication units that are sharable within a small cloud (which we define as the virtual node). PN runs a Network Operating System (NOS) that is capable of performing distributed system operations. This makes it flexible to perform multi tasks in a virtual mode.

Virtual Nodes (VN's) are the core unit of the proposed architecture, where they can connect to each other via Point-to-Point (P2P) MIMO wireless connection and they connect to User Terminals (UT's) via Point - to - Multiple Point (P2MP) MIMO connection. The two type of connection between virtual nodes is Software-based Configurable. VN can serve as a Virtual Base Station (VBS) or Virtual Mobile Switching Center (VMSC) that can connect multiple VBS to Network Resources.

Network Resources are separated into two categories, where delay sensitive resources or Virtual Local Resources (VLR) includes (Signal processing, Radio controllers, Packet Switching, cache memory, etc…), are distributed among edge VN's and resources that are not delay sensitive or Virtual Remote Resources (VRR) includes (SDN, Network Intelligence, HSS, etc…), are centralized virtually into a Cloud Resources Center (CRC) or Network Head.

Figure 1 presents an abstract logical view of the SMMIMO Network Infrastructure.

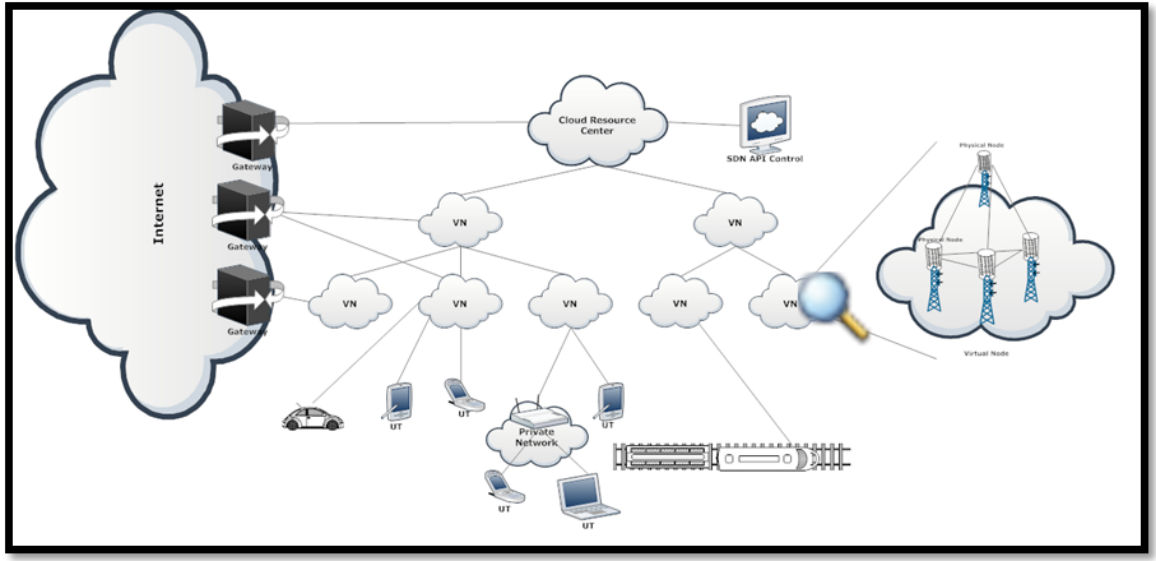

**FIGURE 1 SMART MASSIVE MIMO INFRASTRUCTURE**

The proposed Infrastructure offered a two level virtualization. At the first layer, virtual P2P MIMO connection allows the construction of backbone network links. This type can serve as permanent dedicated Leased Lines, in another hand, P2MP MIMO virtual links are capable of creating Virtual LAN's. The combination P2P and P2MP can construct any desired virtual Network. At the second layer of virtualization, Virtual Permanent Networks of the first layer can further be used to create virtual Networks using SDN API.

The two layer virtualization make it possible to for SMMIMO Network to afford IaaS (at first layer) and NaaS (at the second layer). As a possible scenario of the application on this features, SMMIMO Service Network Provider (NSP) can afford (at first virtualization layer) a permanent Infrastructure Network Service (IaaS) for Network Operators (NO's) and those can afford (at second virtualization layer) Virtual Network Services (NaaS) for their Clients.

Multiple Distributed Internet Gateways (MDIG) is proposed to offload the network from Internet data packets and leave the network backbone ideal to trunk different virtual network packets.

The flexibility of this network and the multi-purpose nodes will eventually decrease the service cost and optimize the exploitation of network global capacity.

Other features like Energy Harvesting Antenna Array (EHAA) and LED Screen Antenna Array (LSAA) can be implemented to decrease the energy cost and optimize the scenery of future smart cities.

## 1.4   Paper Organization

This paper is organized as follows:

Section 2: Introduces the core of the proposed SMMIMO infrastructure.

## 2 NETWORK INFRASTRUCTURE

### 2.1 Infrastructure Overview

SMMIMO combine recent technologies, includes (SDN, NFV, Cloud Theory, Cooperative and Joint Beamforming), in order to afford 5G demands. In addition Distributed Gateways, Parallel Processing, Grid Computing, and Distributed Systems make SMMIMO infrastructure a proposed candidate to build the future Network.

The Network Head Cloud (CRC) is presented by a Cloud Computing Environment capable of monitoring and controlling network infrastructure and includes many long stored information as HSS and Network intelligence (NI) data.

Virtual cloud nodes are connected in the hierarchal model to form the overall network. Each Virtual Node (VN) is composed of distributed resources that make it capable of performing signal processing and computational issues. VN's try as much as possible to reduce the upward data flow by performing data compression and other decision intelligence issues. The VN is the responsible unit to serve its child nodes on their demand.

The overall Network Hierarchy can be extended by new child virtual or physical node. This makes the network flexible to be extended from any region and allows the private local networks to form their own nodes and perform several Device to Device (D2D) communication.

### 2.2 Physical General Purpose Nodes

Physical Nodes (PN) are Radio Access Heads composed of computational and communication units.

PN can be seen as classical massive MIMO base station, but they are extended with several features that make them flexible and capable of performing cooperative tasks. Each PN can be built by assembling several blocks of computational and communication modules (CCM). CCM is a block contain data processing unit, a memory unit, storage unit, Expansion Slot, Radio transceiver unit (TX/RX) surrounded by an array of small size antennas. This allows PN's to be assembled with the flexible size of resources. To build a larger PN, you have just need to add more CCM blocks (taking into consideration technology limits) (see fig2).

Exploiting recent technologies on parallel processing [5], coped with Grid Computing (GC) [6], [7]

Each PN runs a Network Operating System (NOS), that is capable of affording distributed system environment and virtualization services.

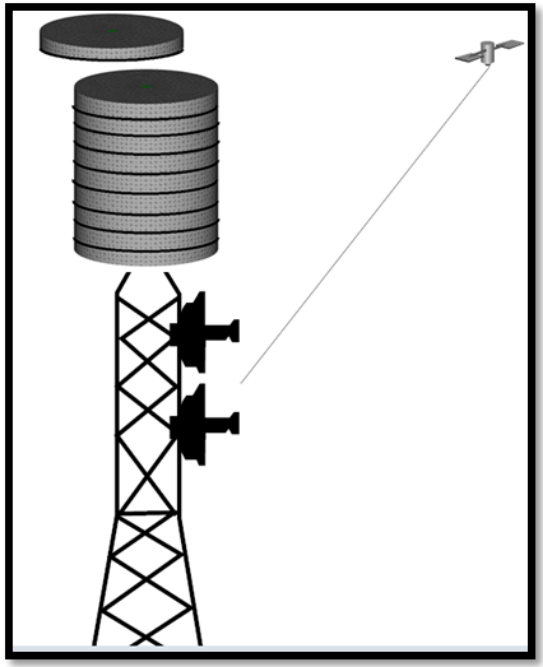

**FIGURE 2 PHYSICAL NODE**

## 2.3 Virtual Nodes

VN's are a cloud of resources selected from multiple PN. On the same manner, the resources at each PN can be distributed on many VN's (fig3, fig4).

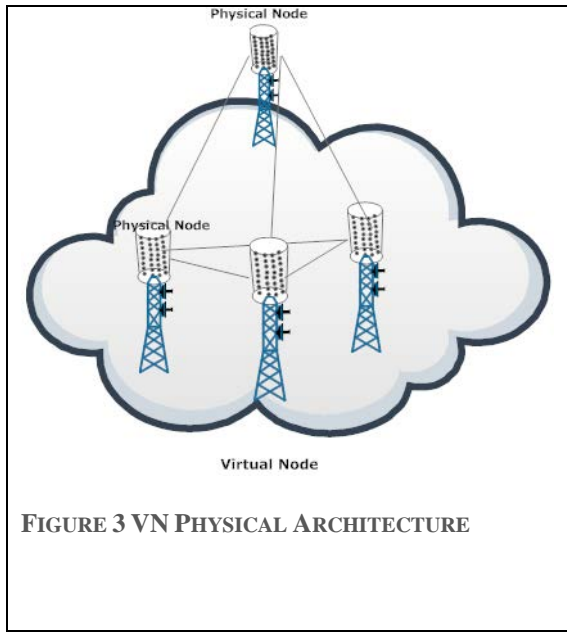

**FIGURE 3 VN PHYSICAL ARCHITECTURE**

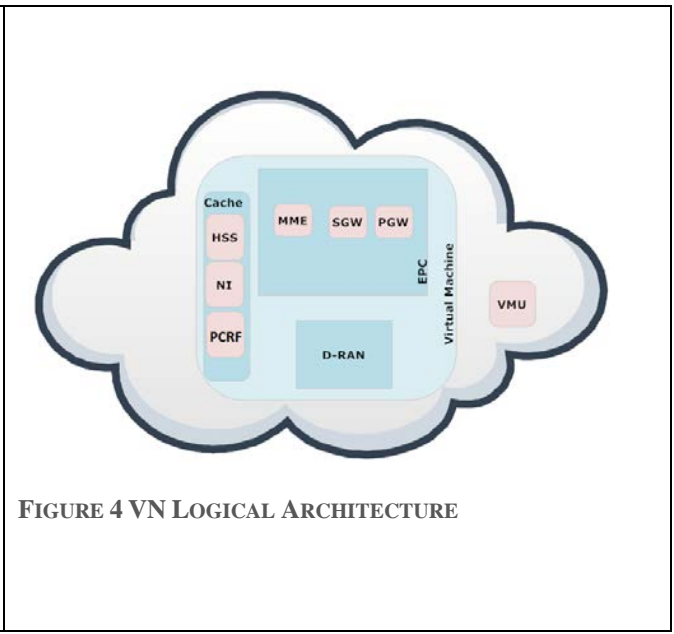

**FIGURE 4 VN LOGICAL ARCHITECTURE**

- Cooperative Distributed Systems

    The cooperation of resources from different PN's makes VN a powerful cloud capable of performing high order processing, storage, and communication tasks. A group of NOS's ($NOS_1$, $NOS_2$, …, $NOS_n$) distributed on PN's ($PN_1$, $PN_2$, …, $PN_n$) collaborate with each other to run a Virtual Machine that is responsible for holding VN task. With such technology, VN can run any type of computational and communication tasks. VN can serve as a Classical mobile switching Center (MSC) and as Radio Access Head (RAH) at the same time. Flexible resource management and task management is driven by SDN API makes it possible to afford Network Virtual Features to local and descendant nodes.

- Distributed Radio Access Network (D-RAN)
    Joint Beamforming with Virtual Antenna Array offered a flexible technique to group distributed antennas from different PN to serve as a Virtual Base Station (VBS). In this paper, we define this VBS as a D-RAN. We define the region covered by the D-RAN as the Virtual Cell (VC). It's clear that VC is the common coverage region of the group of antennas distributes on multiple PN. The coverage region of each antenna can be learned by several quality satisfaction feedbacks of UT's. This makes UT's has the main impact on how the shape of the VC should be.
    Due to the geographical distribution of antennas and the role of UT's to select their serving antennas, system capacity, and coverage quality will increase dramatically in order to obtain better QoS. Other issues related to interference isolation will be discussed later in this paper.

- Evolved Packet Core (EPC)
    EVC is distributed on all VN's, which will lead to reducing backhaul signaling and End to End (E2E) delay. The distribution of Packet Gateway (PGW) into all VN's will reduce the flood of data traverse the

network core and enhance the QoS. This claim is logical since more than 75% of the data that traverse the network are delivered to the Internet network.

Mobility Management Entity (MME) will have direct fast access to each User Terminal (UT) and with the aid of NI unit it will be able to predict UT future positions.

Switching GateWay (SGW) can cooperate with upper VN SGW to quickly switch any connection to its destination.

The distributed EPC proves its ability to reduce E2E delay, backhaul overhead, and QoS.

With the virtual and software-defined environment, EPC functionality can be enhanced and enlarged on parent VN's and reduced in child VN's as needed.

- Virtualization Management Unit (VMU)

  This unit is responsible for managing virtual machine resources including computational and communication resources another function of this unit is to manage the mapping between Physical and virtual environment. This unit is the interface of the virtual machine to the physical Environment and to the SDN API.

- Cached HSS, NI, and PCRF

Policy Control and Charging Role Function (PCRF), Home Subscriber Services (HSS) and Network Intelligence (NI) information are already centralized resources in the CRC network head cloud, but in order to reduce the flood of U-plane and C-plane packets in the network backbone and to reduce End to End (E2E) delay, a cached version related to the local VN is updated on demand to serve end user devices.

The VN can be seen as a small complete network infrastructure cloud capable of communicating efficiently with other VN's. The flexibility acquired by virtualization, make distributed VN's capable of affording any kind of network services.

## 2.4   User Terminal Centric and Virtual Cell

SMMIMO introduces a UT-centric model, where coverage regions depend on the UT position and feedback. On this manner, one of the possible Radio Access Methods that can hold is defined as the Delay Based Map (DBM).

DBM is a mapping technique that defines the UT positions based on the delay distance from each antenna. This can be done on fully synchronized systems, where all devices in the system are synchronized to the same clock.

DBM works as following:

1- UT demand a connection to the system by broadcasting a well-known pilot.
2- Each antenna in the system that can efficiently receive the pilot forward the arrival time stamp and the Signal Quality Weight (SQW) with the antenna ID to his parent node.
3- The Parent node receives many versions about the UT pilot from competing antennas.
4- Parent node then decides which set of antennas can best serve the responsible of this pilot.
5- Each granted antenna had to update the delay distance and the SQW to its own DBM.

After several learning process, where no new major updates will take place, each antenna will have his own DBM. The position of any new UT in the territory can be identified by the delay distance intersection on the DBM of multiple antennae. The intersection region of the group of DBM will form what we define as a Virtual Cell (VC) of the related antenna group.

By DBM technique interference isolation will be easily managed and any new terminal on the map can perfectly define for which virtual antenna array it belongs.

The huge amount of information for each antenna need a large amount of space to be stored, but with the Cloud Computing technology this issue was solved.

All DBM's can be saved in the CRC and only a cached version related to current system state will be temporarily stored in the VN.

Using DBM a lot of Network Intelligence (NI) tasks can be done, includes (UT tracking, Roaming Management, Handover Management, Uplink Pilot Interference Isolation, VC Interference Mitigation, VN Resource Management, etc…).

Because of the virtual nature of the system features, techniques other than DBM can be implemented, taking into consideration the physical Channel Stat Information (CSI) characteristics as a primary factor to define the system model.

### 2.5 Point to Point and Multi-point to Multi-Point Links

The massive amount of antennas and the flexible computational resources in the PN make it possible to assign a group of antennas to serve as a virtual endpoint interface. This interface can be deployed to be connected to another group located on the remote node. This type of connection between nodes is defined by a P2P MIMO link. The flexibility of this link to combine the variable amount of communication resources will turn this link to growing into a backbone Network link or to be limited to connect private indoor node.

Another kind of links can be offered By SMMIMO is the P2MP links where the many-to-many relation between terminal nodes can take place. P2MP can be defined as a group of antenna distributed within a one VN connected to another group of UT Antennas distributed within the VC territory.

At network end VN's, P2MP links is the most favorable scenario, where moving upward in the network hierarchy, P2P links are most probable to occur.

Because SMMIMO network is software defined, P2P and P2MP can be a formed as a Network Functions Virtual (NFV) which make it possible to afford NaaS.

### 2.6 Network Hierarchy

- Semi Distributive Resources

SMMIMO are not based on centralized resources where service will be prone to delay and bottleneck challenges. In other hand deploying fully distributed resources on each node will increase the cost of service in the dimension of energy consuming and resources redundancy cost.

SMMIMO use a semi-distributed resources scheme that is defined in the following manner:

- Resources that are frequently used and delay sensitive are distributed among VC's includes (EPC, RHH)
- Long term resources are cached into VC on demand.

- Multi-PGW to The Internet

To offload network backhaul core from packet data traffic, SMMIMO proposes to use Multi-Packet Gateway (Multi- PGW). PGW are distributed in all VN's, where small VN's can use their parent VN PGW if necessary. Distributed PGW will increase the overall system capacity by offloading the network core for other services network services.

- General Purpose VN

The fact that VN are occurs virtually and with sufficient amount of resources, we can consider it as a General Purpose VN. By defining the tasks of the VN using software API, one can deploy VN as an MSC node or RAN BS or even both. By applying self-assembly algorithms to the VN we can arrive on a small robust Network Cloud. This Network Cloud can be recruited to serve in any position in the network hierarchy.

## 3    *SDN AND NFV INTEGRATION*

Software Defined Networking (SDN) enables the remote management of data plane to be done by third-party software. The flexibility offered by SDN can be more extended by implementing Network Functions Virtualization. Where Network Service Providers can easily virtualize most of the network features using cloud-based API. More services like NaaS and IaaS can be also provided using SDN application.

The integration of SDN and NFV into SMMIMO is based on several steps starting from physical integration until reaching the most virtual top level.

As previously mentioned, PN can be considered as a general-purpose computer device with an array of radio modules as an Input/output Interfaces. (See fig 5)

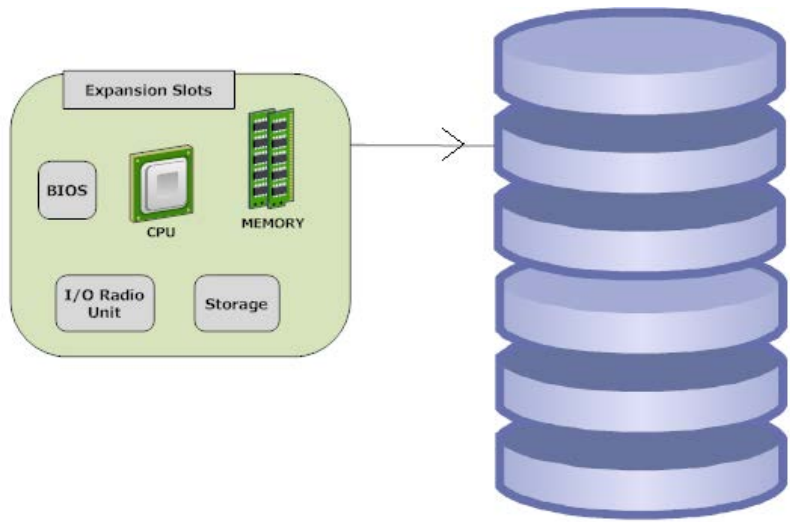

**FIGURE 5 PHYSICAL NODE ARCHITECTURE**

Each CCM block contains CPU, Memory module, a Storage module, an array of I/O radio modules, Basic Input-output System BIOS and Expansion slot.

By assembling multiple CCM with each other we will arrive on a large CCM unit that will present the PN.

The First CCM block will be considered as the master block where it's integrated BIOS will monitor all resource addresses on the PN, and is responsible for starting the NOS deployed on the assembled Storage modules. The overall PN system architecture can be seen as a pool of parallel connected resources. (see fig 6)

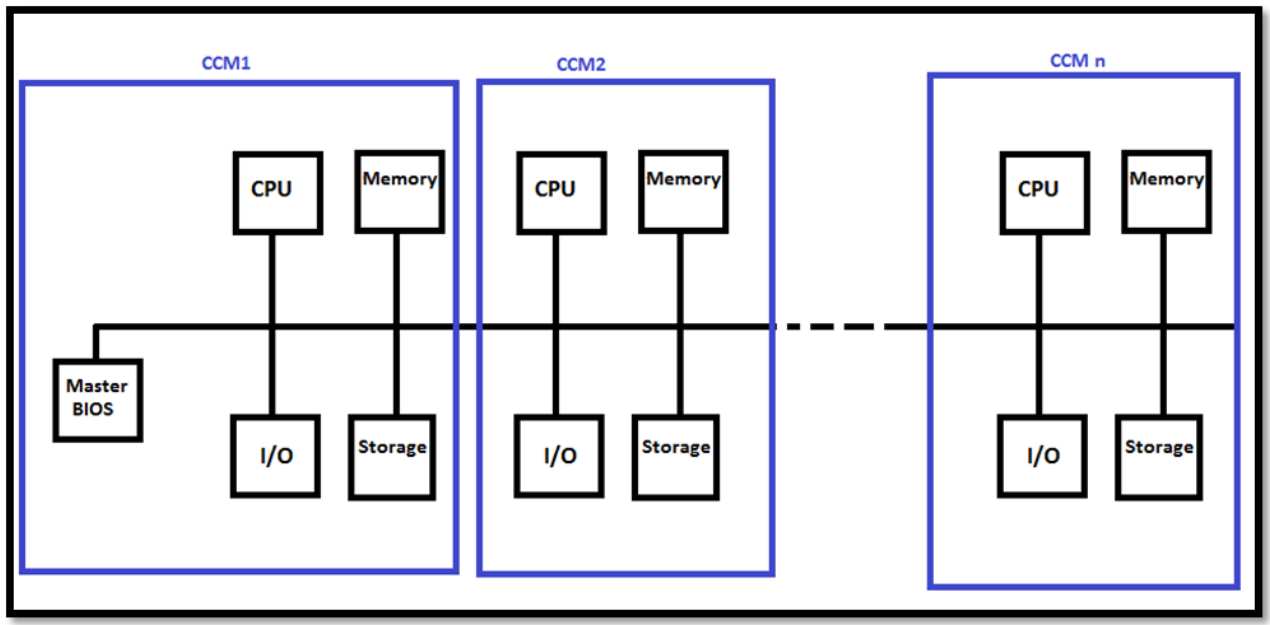

**FIGURE 6 BLOCK DIAGRAM OF CCM POOL**

The system initializes on different stages, where each BIOS starts its own Power on Self-Test (POST) program then report its results to the master BIOS. All connected resources configuration and addresses are monitored by the master BIOS.

Stage 1: After all BIOS's Checkup are successively finished, master BIOS will run the NOS from the boot Storage address.

Stage 2: NOS Checkup the connected resources on the local PN and run a routing and pathfinder algorithm that will discover neighbor PN and manage a basic connection to reach all nodes.by finding the best path using class1 messaging. After Stage 2 is done all the PN will be connected to each other by basic links.

Stage 3: NOS run Virtual Management Service to enable SDN API located at the Network Head Cloud to send Class 2 messages in order to manage Distributed NOS System Clouds.

Stage 4: Distributed NOS, formed by Several PN will then create the Virtual Machine (VM) that will run the VN Cloud. (See fig 7)

VN functions are defined by a set of algorithms run on a Virtual Machine. The Virtualization Management Unit (VMU) will hold all the mapping between Physical and Logical Resources and will serve as the management agent for the SDN API.

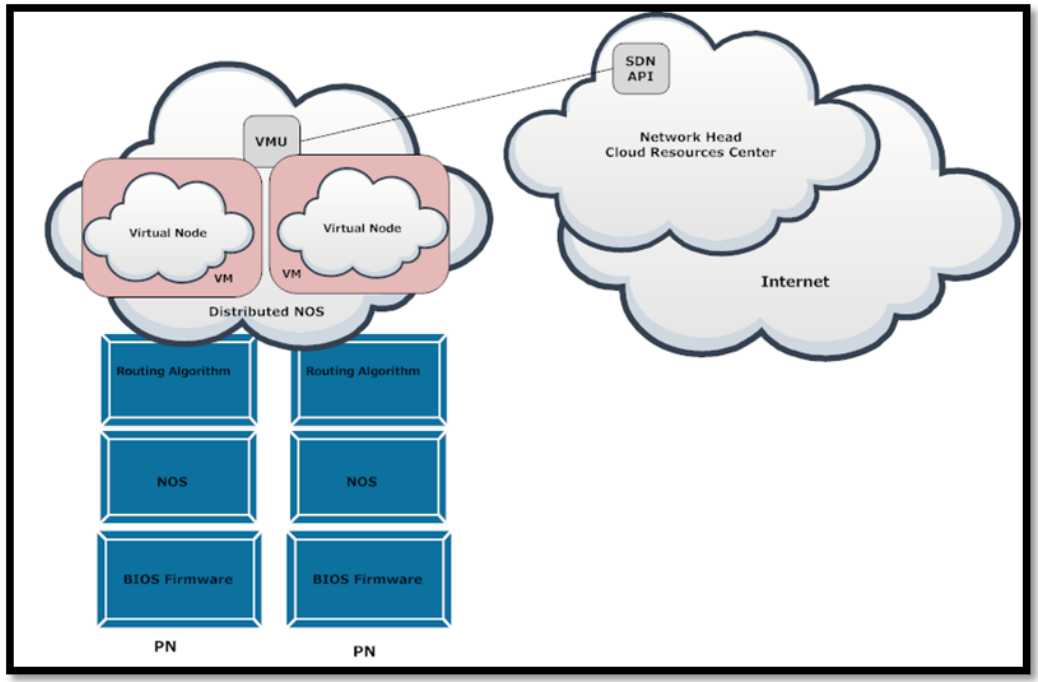

**FIGURE 7 LOGICAL SYSTEM INITIALIZATION**

The Virtual Node on the top of a Virtual Machine (VM) will run by distributed NOS's.

Using the SDN API SMMIMO can provide NaaS, IaaS with simple software configuration. The virtualization of the network functions makes it possible to change the network backbone and Infrastructure without changing the physical platforms.

Recent researches on Distributed Systems prove their ability to run cooperative tasks even when physical platforms are widely spread. An application example of the architecture of Distributed Systems that can run on advanced programmable networks is presented by InstaGENI [8]. This software architecture is designed to provide deeply configurable and deeply-programmable IaaS and customizable OpenFlow NaaS.

## 4    VIRTUALIZATION LATENCY

The high order virtualization in our model seems that a large gap between physical and logical environment had been done, which will lead to virtualization latency in order to reach the physical environment.

A solution to this challenge had been proposed by [9]. A novel approach to minimize the virtualization overhead for low latency network for High-Performance Computing Clouds. Performance results of implementation over a multi-technology software defined network are promising. The efficiency of the proposed low-latency SDN is analyzed and evaluated for high-performance applications. Experiments were done on Mellanox FDR InfiniBand interconnect and Mellanox OpenStack plugin shows the best performance for implementing VM-based high-performance clouds with large message sizes.

Virtualization cost is negligible compared to the system optimization granted by this technology. Other optimization on the virtualization latency reduction can be done by grouping cooperated resources based on the Delay Access Time (DAT).

# 5    COST-EFFECTIVE NETWORK

SMMIMO designed to be aware of providing low-cost services, in this manner we will address two bold features that make SMMIMO considered as a 5G cost-effective network.

## 5.1    universal Service Quanta Unit

The diversity of metrics that affect the QoS and the engage of new services with different demands on quality leads to thinking of new service unit that includes much more quality metrics.

Where pricing based on Time-of-Use, Bandwidth and bit transferred are not sufficient to price cost –effective services.

For example, many network operators ignore the cost of control signaling messages (e.g. call setup messages) and afford unlimited signaling messages for free. With nowadays smartphones and by applying a small part of the code to the system, one can flood the network with false signaling. This attack will be promoted to a Denial of Service (DoS) attack by spreading the code on many smartphones. With the help of application Stores, thousands of versions of this codes can be installed quickly.

Another type of attack that can be exploited by the free signaling commands can be done by encoding a set of the signaling message to send free text messages between devices. The sender application encodes the text message at the sending device and runs it as a series of signaling codes that can be decoded at the second end to be converted back to the original text.

These and more attacks can be done on signaling messages that can reduce the overall performance of the network.

We propose a new Service Quanta Unit (SQU) that takes into consideration all service costs. Network service providers, network operators, and end users can be priced based on this universal unit.

SQU must include many additional metrics in order to afford fair pricing and cost-effective services.

The following are some additional proposed Metrics:

- **Data Urgency Level**:  addresses the demand on messages sensitivity to delay.
- **Energy cost:** addresses the energy cost of the data unit to reach the destination.
- **Distance to Destination:** addresses the occupation of network elements of the data unit to reach its destination (e.g. number of hops, backbone distance travel, etc…).

- **Signaling Cost:** addresses the cost related to network and user signaling tasks.
- **The quality of Content:** addresses the demanded on the quality level of the content that are delivered (e.g. Multimedia content and Over The Top (OTT) players packets can be encoded and compressed by lossy techniques where text messages may require lossless techniques).

Other metric can be added to form a standard universal SQU that can improve network services and leads toward fair and cost-effective services.

Network operators and middle agents will compete to afford a low SQU pricing to the end users by optimizing their virtual network Infrastructure. It is expected that this approach will optimize the overall network performance and reduce the service cost and the 5G demands on QoE will be solved.

### 5.2 Pay-as-You-Go and Go-as-Your-Size

The use of SQU simplify the service pricing and enable fair service, where service provider and clients of any size can deal.

By virtualizing resources, SMMIMO can offer a flexible size of services which will fit any organization size. The client can have basic services (e.g. IOT services), up to Infrastructure services (e.g. network operators). We call this flexibility "Go-As-You-Are (GAYA)", where any size of clients can have a service that fits him.

PAYG, GAYA will afford a cost-effective, low-cost high-quality services. While using General purpose PN will decrease the cost of Network Devices. These approaches meets the key challenges of most 5G ongoing projects (e.g. [10], [11]).

## 6  ENERGY-EFFICIENT NETWORK

Massive MIMO technology is based on focusing the beam into a specific space, which allows spatial multiplexing as a new multiplexing dimension. This technique proves its ability to reduce transmission Energy relative to throughput. Moreover, virtualization technology will reduce energy consumption due to the reduction of hardware units and the use of software instead.

5G need to be a Green Network by increasing Energy Efficiency (EE) and using clean sources.

Many ongoing types of research are working on energy harvesting communication networks, where authors in [12] shed the light on recent advances in Energy Harvesting Wireless Communications (EHWC).

We propose the use of grid solar cells that can be stacked on the antennas array of the PN as an Energy Harvesting Antenna Array (EHAA). The multi-layer module can work as an RF antenna and at the same time as a harvesting solar cell.

Using fractal antenna technology the antenna array can displace a small space on the CCM which allows a large number of antennas to be implemented on the PN.

An example of such a multi-layer antenna was introduced in [13], where an aspect of their disclosure is directed to surfaces that include dual-use or multiple-use apertures. As they mention "Such aperture engine surfaces can include a top layer of antenna arrays, a middle layer of a metal-fractal backplane player, and a third (or bottom) layer for a solar cell or solar oriented power collection".

By combining Energy harvesting technology to the Antenna array of Each PN, service cost will eventually decrease and future smart cities will use a Green Network.

## 7    LEGO-BUILDING NETWORK

SMMIMO introduces flexibility on all layers of the technology starting from flexible hardware implemented on the PN ending with virtual network functions installed as a third-party application on the top of Virtual Machine's (VM's).

PN can be built out of blocks of CCM, NOS's from multiple blocks can be exploited to build a distributed system cloud and this cloud is able to construct multi-virtual Machines to run VN's.

VN hold major network features include virtual switching, Logical Interfaces, and MME. Changing roles between network features or even add new ones can be easily done by updating the running software.

Furthermore, SMMIMO proposes a self-assembly VN. This feature can be installed as a software on the top of NOS's running on different PN's.

The self-assembly algorithm takes place at stage 3 of system initialization process. Based on previous knowledge about the antenna occupation regions, self-assembly algorithm can group resources to form VN's automatically without the need of SDN Controller supervision. This can be done using the basic network that is managed at stage2.

Self-Assembly Algorithm works in the following manner:

1- Broadcast echo message to discover neighbors
2- Using best path selection algorithm an initial connection map to neighbor PN will be done.
3- Discover neighbors map to update and optimize current connection map to all destinations.
4- At the end, a connection scheme to all PN in the Network is ready to connect VN's.

Self-Assembly Algorithm can be used on system failure and catastrophes to ensure system robustness and network availability.

## 8    HETEROGENEOUS MULTI-PURPOSE NETWORK

The services of SMMIMO infrastructure afford a comfortable environment to deploy the different type of networks virtually within the same Infrastructure. A client of different demand size can use this infrastructure.

### *8.1    Private Networks*

Enterprises and companies of large size can use their own virtual network, point-to-point links, permanent leased lines, Packet Switched Network (PSN) and Circuit Switching Network (CSN).

Network Operators can also benefit from IaaS to form their own Networks.

*8.2 Indoor Networks*

The node hierarchy architecture makes it easy to extend the network with private small nodes. Private nodes with limited resources will serve as a private network extension Access Points (AP's) that can be used indoor. This will bring a small BS to the indoor environment to increase coverage and reduce service cost. Using Mobile station as a terminal node, ad-hoc network, and D2D communication will be easily managed.

*8.3 IOT Network*

Internet of Things (IOT) consumes a lot of network resources, where each thing need to manage a connection to the internet network. These connections do not consume a lot of bandwidths but, a lot of connections and signaling will reduce the network performance and will flood the network backhaul with overhead signaling messages.

Many features proposed by SMMIMO infrastructure that helps to reduce this challenge. The flexibility of SMMIMO is also applying to QoS, where service is quantized into small clearly defined scalars. This flexibility allows the use of basic connections that can handle a large amount of devices connected to the VN with optimal resource allocation.

Another feature of SMMIMO is that most VN's are connected to Packet Gateways (PGW) which will reduce message signaling on the network backhaul by offload the network from internet Packets.

## 9 NETWORK CLOUDING AND SERVICES

SMMIMO can be seen as a large network cloud of servers equipped with a massive amount of radio access devices. The general purpose computation cloud allows Any-Thing-as-a-Service (XaaS). In other words, the virtualization of physical things can make them affordable as a service for clients. By focusing on communication services, SMMIMO can serve as an interface for public Clouds located within the Internet.

## 10 NETWORK INTELLIGENCE

Adding Artificial Intelligence (AI) to Information and Communication Technologies (ICT) brings interesting advances in the domain of scalability, Network Security, Network Monitoring, Energy Management, Bandwidth Management and other important fields.

The two major implementation of AI where applied on:

1- Cloud Computing and IOT
2- Network Security and Network Management

SMMIMO address the two preceding points and the need for to apply AI on this network is considered the main issue.

The integration of the NI unit at the CRC (where there exist of Data Center and Computational resources), facilitate the task to applying Deep Learning Algorithms and Decision Making Algorithms ,taking into advantage the huge amount of data stored at the CRC. Such algorithms will have a great optimization impact on many challenging fields on the future Network.

- **Network Security:** There are a lot of advances made so far in the field of applying AI techniques for combating cyber crimes [14]. Intrusion Detection and Prevention System (IDPS) can be applied as a software that uses ontology and knowledge bases to prevent the network from criminal's threat and run real-time attack detection. Artificial Immune System (AIS) applications also can run on the NI unit to uphold stability in a changing environment. AIS mimic the behavior of the biological immune system and apply intelligent techniques to protect the network.

- **Network Management:** The flexibility offered by SMMIMO can be exploited in an intelligent technique to manage an equilibrium balancing technique to dynamically move network capacity from less congested areas toward the jammed area. This can be done by system monitoring and learn through several epochs. For example city center and industrial areas demands high bandwidth at the working hours while residential areas increase its demand on bandwidth at evening. The ability of the network to learn from previous patterns using AI Algorithms coped with the ability of the network to dynamically modify its structure will optimize the exploitation of network bandwidth without additional costs.

- **Mobility Management:** SMMIMO infrastructure uses the user terminal physical characteristics (e.g position and delay from antennas) to isolate interference and make easy channel reuse. This property can be exploited to predict the future channel characteristics of the UT based on previously learned state. For example using the delay distance map (DBM), NI unit can monitor the Mobility of the UT and predict the next position of the UT based on the motion vector and speed of this terminal. This will allow NI unit to propose the next group of antenna that will serve a specific UT .

The Softwarization of the network equipment allows many AI techniques to be applied in order to enhance the network performance.

## 11   NEW BUSINESS MODEL

### 11.1   Opex- Oriented business model

SDN and NFV offered by SMMIMO will change the business model from CopEx-Oriented to OpEx-Oriented. The evolution from Hardware to Softwarization can be expected to have a big influence on the business equation [15].

Network programmability enabled by SDN, allows network functions to be afforded in a dynamic way. This will have a positive impact on service providers business by boosting revenues and decrease the total cost of ownership that depends basically on (OpEx and CopEx).

The capital and operational expenses are supposed to be decreased dramatically, where operational Expenses (OpEx) will be the basic player in the new business equation.

Software developers and service brokers will be major players in the new business model.

### 11.2   Service Trading and Network Sharing

To cope with the demands on QoS network operators had to afford a complete coverage on all the geographical territory under their service. But this issue depletes more resources on low populated areas. This problem accompanied mobile networks for a long time ago, but the virtual fashion of resources and with clear defined SQU, this problem can be contaminated by service sharing.

VN's can rapidly create a point to point link between two virtual networks to enable smooth handover coverage without creating a new connection. This can be done by service trading and sharing between two network operators. The sharing of logical resources will lead to a higher order of network exploitation with lower service cost.

Service trading will be a major career for all size of clients, (e.g. one can think of distributing virtual networks on remote and less populated areas and sell on-demand services to network operators).

The OpEx oriented Business model coped with service trading will extend the market with new carriers which will help reduce unemployment in the new smart cities.

## 12   CITY SCENERY AWARE DESIGN

An important issue that must be taken into consideration when brings new technology to public places is the scenery impact on the city. Implementing large poles, each of them holding thousands of antennas into the city centers will distort the city outlook.

As our design takes into consideration the outlook of the PN, we add two propositions that will improve the outlook appearance:

- **EHAA:** The solar cell grid layer that contains a black glass external layer is the best choice for Energy Efficiency and PN Scenery in the daytime.
- **LSAA:** as the PN holds a large space of grids (antenna and solar cell layers), we propose to add the third layer of LED grid array. This array can serve as a big screen that runs advertisements and public videos, which has a positive impact on the city scenery night view.

With such a design future, smart cities will look better and without reducing the network performance.

## 13   PRIMARY RESULTS

To test the capacity of VC we introduce the following scenario:

For simplicity, we consider a group of 4 PN's communicate with each other to form 4 VN's. Each VN has 1000 antenna and forms a virtual cell that holds 100 UT antennas. The virtual nodes grouping is supposed to be formed based on the judgment of another upper VC. The parent VC is supposed to receive forwarded pilots from different antennas distributed on the 4 PN's. We further suppose a normal distribution of UT's which leads to form an equal resources VC's.

The simulation scenario is done on a UT position centric model similar to that introduced in section 2, where a Delay Based Map (DBM) is created to isolate interference.

The introduced results correspond to one VC, where $\alpha \in (0,1]$ is the interference factor and $\mu \in (0,1]$ is the minimum acceptance ratio of received power that entitles the received antenna to be one of the candidates competing to serve the requesting UT (see fig 8).

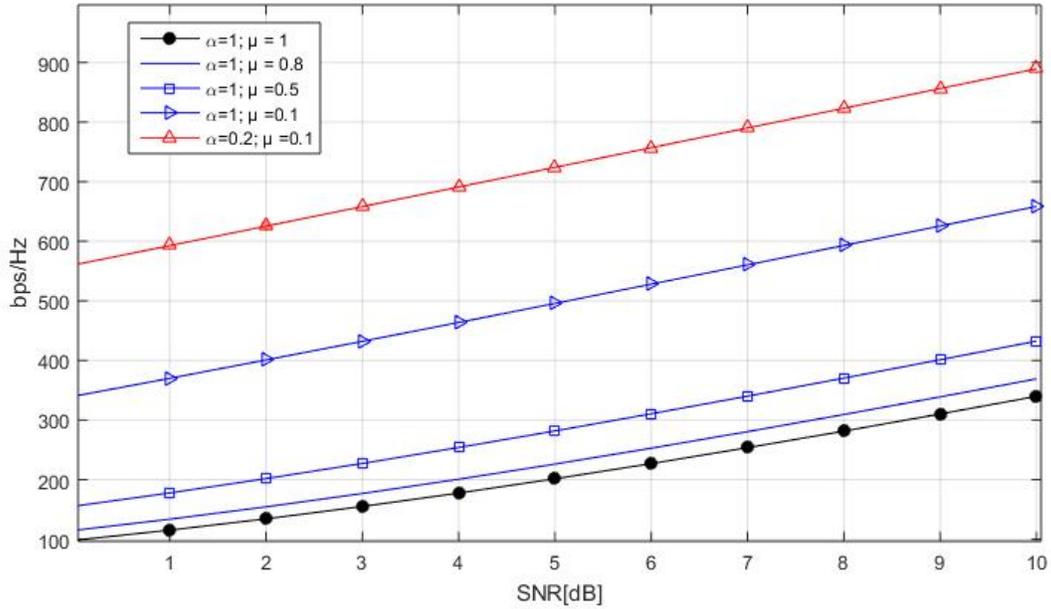

**FIGURE 8 ERGODIC CAPACITY VS SNR**

The lowest curve with $\alpha = 1, \mu = 1$ matches a classical MIMO system with full interference and no threshold acceptance value.

Threshold value $\mu$ can be assigned by network engineers and $\alpha$ can be reduced by several un supervised learning process of the system.

The Ergodic capacity of the VC reaches about 900 bps/Hz at SNR of 10 dB, which is a promising result for implementing VC in 5G Networks.

## 14   CONCLUSION

SMMIMO introduce a new view for 5G networks that encourage researchers working in this field to add innovative ideas to their Infrastructures. The Cloud Networking with high flexibility takes 5G into a new horizon beyond the classical hardware limitations.

SMMIMO is a collection of Latest technologies on clouding and communication theories and applications. The primary results show that we are not far from reaching the desired 5G Network.

As a communication issue, joint Beamforming with Virtual Cell approach proved its ability to increase the system capacity, while cloud computing and virtualization outrun the experimental stages and are ready to be applied on the new Clod Networking.

As we think, the 6G Infrastructure may come on as a software to be installed on the systems via the head cloud. We propose to increase the overall system performance by improving the virtualization impact. Future researches on virtualization overhead and visualization risk management will be very helpful for the future network infrastructure.

components". United States of America Patent US Patent 9,035,849, 19 May 2015.

[14] S. Dilek, H. Çakir and M. Aydin, "Applications of Artificial Intelligence Techniques to Combating Cyber Crimes: A Review," arXiv preprint arXiv:1502.03552, 2015.

[15] E. Hernandez-Valencia, S. Izzo and B. Polonsky, "How will NFV/SDN transform service provider opex?," Network, IEEE, vol. 29, no. 3, pp. 60-67, 2015.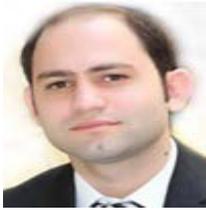

**Biographical Notes:**

1Ahmed A. Abboud (2nd Oct 1985) was born in MazratMichref South Lebanon. Received a technical diploma in communication & computer engineering from University Institute of Technology (UIT) Jwaya, South Lebanon and the Master of Science in Computer Science & Communication from Arts, Sciences & Technology University in Lebanon, in 2008 and 2012 respectively. He is now a Ph.D. student at the University of Limoges since October 2014.His research interests are in the area of applying artificial intelligence algorithms on MIMO communication systems.

19